# Mapping Patent Classifications: Portfolio and Statistical Analysis, and the Comparison of Strengths and Weaknesses

Loet Leydesdorff,*[a] Dieter Franz Kogler,[b] & Bowen Yan [c]

**Abstract**

The Cooperative Patent Classifications (CPC) recently developed cooperatively by the European and US Patent Offices provide a new basis for mapping patents and portfolio analysis. CPC replaces International Patent Classifications (IPC) of the World Intellectual Property Organization (WIPO). In this study, we update our routines previously based on IPC for CPC and use the occasion for rethinking various parameter choices. The new maps are significantly different from the previous ones, although this may not always be obvious on visual inspection. We provide nested maps online and a routine for generating portfolio overlays on the maps; a new tool is provided for "difference maps" between patent portfolios of organizations or firms. This is illustrated by comparing the portfolios of patents granted to two competing firms—Novartis and MSD—in 2016. Furthermore, the data is organized for the purpose of statistical analysis.

[a] * corresponding author;
University of Amsterdam, Amsterdam School of Communication Research (ASCoR), PO Box 15793, 1001 NG Amsterdam, The Netherlands; email: loet@leydesdorff.net.
[b] School of Architecture, Planning & Environmental Policy & UCD Centre for Spatial Dynamics, University College Dublin (Richview Campus) Belfield, Dublin 4 D04 V1W8, Ireland; dieter.kogler@ucd.ie.
[c] SUTD-MIT International Design Centre, Singapore University of Technology and Design, Singapore 487372; e-mail: bowen_yan@sutd.edu.sg.

## 1. Introduction

Patent data provide a primary data source for scholars interested in the development of technological knowledge (e.g., Strumsky *et al*., 2012). In a comprehensive study of US patent data, Jaffe and Trajtenberg (2002), for example, argued that patents and patent citations provide "a window on the knowledge economy." However, patents are indicators of inventions and not innovations (Archibugi & Pianta, 1992; Pavitt, 1985; Grilliches, 1990). Like publication data, patent data provide a wealth of information about knowledge claims (Hunt, Nguyen, & Rodgers, 2007), references to prior art, forward citation, inventor and applicant addresses, and classifications. Note that patenting can also be strategic or defensive given a firm's technical expertise and product portfolio (Alkemade *et al*., 2015).

Different from journal articles, patents are not refereed among peers, but by examiners at the patent office, among other things, on the criterion of whether the submitted application provides novelty. The purpose of the novelty requirement is to prevent prior art from being patented again. The examiner is entitled to add references to prior art,[1] but most importantly organizes the patent in a patent classification system. In the absence of an equivalent for journals, the classification system can be considered as the intellectual organization of the database of novel products and processes of economic value.

Jaffe (1986) was the first who proposed to use co-classifications of patents for characterizing the technological positions of firms with the objective of quantifying technological opportunities and research spillovers. For this purpose, each patent can be represented as a vector of classifications. Jaffe (1989) used the cosine between these vectors as a similarity measure (Salton & McGill, 1983). Sets of patents (e.g., patent portfolios of firms) can be modeled as aggregates of these vectors into matrices. The resulting matrix can be used for statistical analysis and/or the visualization of the similarities as projections on a map (Schiffman, Reynolds, & Young, 1981; Waltman, van Eck, & Noyons, 2010).

---

[1] At EPO all references are added by the examiner.

In a series of studies, we developed instruments for the mapping and analysis of patent data with a focus mainly on data of the US Patent and Trade Office (USPTO). US patenting is considered the most competitive and is therefore most relevant to technology and innovation studies (Granstrand, 1999; Jaffe, 1989; Jaffe & Trajtenberg, 2002; Lee, 2013). USPTO maintains a freely available database of patent data at http://patft.uspto.gov/netahtml/PTO/search-adv.htm. This database is copied at other places, such as Google Patents, the National Bureau of Economic Research (NBER) in the USA, and PatentsView (http://www.patentsview.com). PatentsView adds to the data by providing disambiguated inventor and assignee identifiers (Monath & McCallum, 2015).

Building on approaches suggested in previous studies (e.g., Breschi *et al*., 2003; Leten *et al*., 2007; Verspagen, 1997), we developed instruments for the mapping and analysis of patent data using International Patent Classifications (IPC). IPC was developed by the World International Property Organization (WIPO) and further elaborated by the European Patent Office (EPO) into the European Classification System ECLA. This classification uses up to fourteen characters for the indexing. As of 1 January 2013, however, USPTO and EPO use the new Cooperative Patent

Classifications (CPC) which build on ECLA. In addition to the various patent offices of the EU member states, China, Korea, Russia, and Mexico, among others, are in the process of aligning their classifications with CPC. In this study, we use CPC for the mapping, review and integrate the results and routines from the previous studies, revise some of the methodological choices, and update the routines and analysis accordingly.

**2. Choices of parameters**

In addition to the choice of a similarity criterion such as the cosine, a number of parameter choices are relevant for the mapping of patent data: the choice of domains in terms of patent offices (USPTO, EPO, WIPO, etc.), the classification system (CPC, IPC, etc.), the clustering algorithm (e.g., co-classification, co-citation, bibliographic coupling), etc. The classifications can be used with different levels of detail by using more digits of the respective classes, i.e. main- and/or sub-classes.

Kogler *et al.* (2017a) used IPC at the four-digit level for portfolio analysis; Yan & Luo (2017) used IPC at the three-digit level (IPC-3). Leydesdorff *et al.* (2014) used both three and four digits of IPC for exploring a dynamic approach. With similar objectives, Kay *et al.* (2014) composed 466 IPC categories at different levels of depth after preprocessing their data from the comprehensive PATSTAT database[2]. Schoen *et al.* (2012) use a database of the Corporate Invention Board (CIB) for the construction of their own classification (Alkemade *et al.*, 2015). Archibugi & Pianta (1992) combined IPC with the Standard Industrial Classification (SIC) using a concordance table. However, the objective of these last authors was not to provide a map or overlays.

---
[2] Further information regarding access to PATSTAT can be found at: https://www.epo.org/searching-for-patents/business/patstat.html

In this study, we use CPC at the four digit level. CPC is similar to IPC in the first four digits, but the classification of individual patents can be changed in the process of reclassification; new classes have also been added. Furthermore, CPC contains new categories classified under "Y" that span different sections of CPC in order to indicate new technological developments such as nanotechnology and technologies for the mitigation of climate change (Scheu *et al.*, 2006; Veefkind, Hurtado-Albir, Angelucci, Karachalios, & Thumm, 2012). Currently, there are nine of these Y-classes. Since these boundary-spanning classes are back-tracked into the database, they introduce new links between previously unrelated classes. As a result, one can expect changes in the networks among the classes. All values in this study are based on CPC at the 4-digit level (CPC-4) instead of IPC-4.

Co-classification is a binary measure at the level of individual patents. Co-citation of classes or bibliographic couplings among them can be considered as more refined measures of the strength of the relationships. Yan & Luo (2017) compared twelve techniques for the mapping of patent data in terms of citation relations, inventive activities, etc., among 121 IPC classes at the 3-digit level. The conclusion was that maps based on Jaccard-normalized bibliographic coupling at the (disaggregated) patent level—"A1" in the classification of methods provided by these authors— significantly outperformed maps based on the cosine values among of aggregated citations among IPC classes (Leydesdorff *et al.*, 2014 and 2015). Actually, Yan & Luo (2017, at p. 435) conclude that "class-to-class cosine similarity among aggregated citations (A2) performed the worst in various analyses, although it is popularly used for constructing network maps of patents." This critique prompted us to reconsider on previous choices of parameters.

In this study, we combine our two approaches: we use bibliographic coupling among CPC-4 classes at the level of patents as individual documents. We first construct the asymmetrical (2-mode) matrix of patents (cited) in the rows versus citing patents aggregated into the 654 CPC classes at the 4-digit level in the columns. The Jaccard index and cosine values are then computed over the 654 columns (Leydesdorff & Vaughan, 2006). The resulting (1-mode) matrices can be utilized as input into the mapping exercise.

For the clustering and subsequent coloring of the resulting map we use a methodology recently developed by Leydesdorff, Bornmann, & Wagner (2017) for journal mapping. VOSviewer (v1.6.5; 28 September 2016) provides both a community-finding algorithm and visualization (Waltman *et al*., 2010). We use this algorithm for generating a hierarchically decomposed set of maps of patent classes. Although the resulting maps are statistical and cannot claim semantic authority, they can serve the heuristics by offering a baseline. Both the Jaccard-based map and the cosine-based one were decomposed into clusters using these statistics. VOSviewer enables us to visualize 654 data points without cluttering the labels on the screen by foregrounding and backgrounding strong and weak presences, respectively. The CPC classes are fractionally counted so that each patent contributes with a value of one.

## 3. Methods

*3.1. Data*

We harvested patent data from 1976 to July of 2016 from USPTO and PatentsView [3] on January 11, 2017. The data set contains 5,175,268 utility patents. Each patent is classified in one or more CPC classes. The definitions of CPC classes are available among other places at https://www.uspto.gov/web/patents/classification/cpc.html. As noted, CPC and IPC are identical in terms of the first four digits. However, IPC contains 630 four-digit classes, whereas CPC distinguishes 705 such classes, of which 654 are currently in use. We use these 654 classes.

*3.2. Distance measures*

The 5M+ patents cite 7,203,533 patents. First, we constructed a 2-mode matrix of the 654 classes—aggregates of the 5M+ citing patents—versus the 7M+ cited patents. This matrix was then used for computing the symmetrical 1-mode matrices of 654 * 654 Jaccard and cosine values, respectively. Eqs. 1 and 2 provide the formulas for calculating Jaccard and cosine similarity between two random variables.

$$Jaccard = \frac{\sum_{i=1}^{n} x_i y_i}{\sum_{i=1}^{n} x_i^2 + \sum_{i=1}^{n} y_i^2 - \sum_{i=1}^{n} x_i y_i} \quad (1)$$

---

[3] Patentsview is available at http://www.patentsview.org/ .

$$Cosine = \frac{\sum_{i=1}^{n} x_i y_i}{\sqrt{\sum_{i=1}^{n} x_i^2} \sqrt{\sum_{i=1}^{n} y_i^2}} \qquad (2)$$

The Jaccard matrix is based on counting the cited patents as binary [A1 in Yan & Luo (2017)],[4] whereas numerical values can be used for computing cosine values: a patent can be cited more than once in a CPC-class of citing patents. Yan & Luo (2017) categorize this cosine based on values at the individual patent level as a third option A3 and conclude that "in some cases of our analysis, A3 is no worse than A1." However, the QAP (Pearson) correlation between the resulting two matrices—diagonals excluded—is 0.786 ($p<.0001$). Thus, one can also expect the two maps to be different. We explore and compare both maps.

The *cosine* can be considered as a proximity measure in the vector space and (1 – *cosine*) thus provides a distance measure. While the corresponding "Jaccard distance" (= 1 – Jaccard) is widely accepted in the literature as well, we note that the Jaccard index in contrast is a relational measure. From the perspective of graph theory, the geodesic would be a better measure for the distance between two nodes in a network (de Nooy, Mrvar, & Batgelj, 2011). As a distance measure, in our opinion, one should therefore preferably use (1 – *cosine*) and not the Jaccard distance.

---

[4] A non-binary equivalent of the Jaccard index is provided by the Tanimoto index (Lipkus, 1999; cf. Salton & McGill, 1983, at pp. 203f.).

*3.3. Clustering*

Both the Jaccard and the cosine matrices were decomposed using the routine decomp.exe (available at http://www.leydesdorff.net/jcr15/program.htm). The maps based on the Jaccard indices are available (*i*) at http://www.leydesdorff.net/cpcmaps for each class (bottom-up) and (*ii*) at http://www.leydesdorff.net/cpcmaps/scope using hierarchical clustering top-down. One has the option either to access the various maps directly in the .jpg format or to webstart a map using VOSviewer. The latter option provides an analyst with follow-up options such as choosing other parameters or exporting the files in formats used by other software applications.[5]

Jaccard and cosine values generated on the basis of the 2-mode matrix of patents versus classes can be very small since these matrices are sparse. During the decomposition, for example, VOSviewer saves files with fewer decimals than the cosine values so that these values are rounded to zero. This inflates the modularity and generates small clusters. In order to avoid this effect, we multiplied all values by 1,000. For the global map, this makes no difference in the case of cosine values and hardly any difference for a map based on Jaccard values; but the larger values improve the decomposition because the additional zeros otherwise increase the number of small clusters. However, the larger values can no longer be used for the distance measurement because they can be larger than one. For this reason, we use the cosine between aggregated citations among classes as the proximity measure in the routines about portfolio management.

---

[5] For the convenience of the user, the symmetrical Jaccard matrix is made available at http://www.leydesdorff.net/cpcmaps/jaccard-sym.csv. Analogously, the cosine-based maps and decompositions are brought online at http://www.leydesdorff.net/cpc_cos and http://www.leydesdorff.net/cpc_cos/scope, respectively. The symmetrical cosine matrix is available at http://www.leydesdorff.net/cpcmaps/cosine-sym.csv.

(This file "cos_cpc.dbf" is available at http://www.leydesdorff.net/cpc_cos/cos_cpc.dbf ; see the Appendix for details.)

*3.4. Portfolios*

We provide a routine (CPC.exe; see the Appendix) for generating overlays on the new map. Overlays on maps can be used for portfolio management by science-policy makers and R&D management (Kogler *et al*., 2017a; Leydesdorff, Heimeriks, & Rotolo, 2016; Rotolo, Rafols, Hopkins, & Leydesdorff, 2017). We add to the portfolio mapping, the option of a "difference map" so that one can visually compare two portfolios in a single map, indicated with two different colors.

Using the routine (made available at http://www.leydesdorff.net/cpc_cos/portfolio/index.htm; see the Appendix for instructions), one can first retrieve a specific patent set using any search string valid in the USPTO search interface. The routine then generates a file vos.txt that can be read by VOSviewer and generate overlays.[6] In order to facilitate comparisons among sets a column variable is added to a file matrix.dbf for each run. This file can be read into programs such as SPSS for statistical analysis. Similar, a row variable is added to the file rao.dbf containing the value of Rao-Stirling diversity Δ (Rao, 1982; Stirling, 2007) and Zhang *et al*.'s (2016) improved measure $^2D^3$ [= 1 / (1 – Δ)] for the sample under study. In our opinion, these diversity measures can be considered as measures of "related variety" (Frenken, Van Oort, & Verburg, 2007), since the disparity is measured in addition to the variety (Rafols & Meyer, 2010). The variety is ecologically related in terms of the disparity among the categories. If the files matrix.dbf and rao.dbf are not already present or were deleted, they are generated *de novo*.

---
[6] VOSviewer is freely available at http://www.vosviewer.com (Van Eck & Waltman, 2010; Waltman, van Eck, & Noyons, 2010).

The routine CPC.exe first asks for a name of the sample (e.g., "Boston") that is used to label the column and row variables added. The routine is technically similar to the one defined in Kogler *et al.* (2017a) for IPC classes, but based on the improved map using CPC-4. The routine CPC2.exe asks for two sets of downloaded patents in order to generate a difference map. Difference maps can be used for comparing portfolios of different units of analysis.

**4. Results**

*4.1. Jaccard or cosine-based maps?*

The global maps based on the Jaccard index (Figure 1) and the cosine (Figure 2) are obviously dissimilar. Whereas we are able to provide the cosine-based map with an interpretation, we find this difficult for the Jaccard-based one. A lexigraphical approach did not work for making a choice, since the headers of CPC classes are dominated by words such as "material," "device," etc. Table 1, for example, shows the ten most frequently used words in the six clusters distinguished in the Jaccard-based matrix (using VOSviewer), whereas Table 2 provides the equivalent information for the nine clusters distinguished in the cosine-based matrix. Neither table is semantically rich.

**Figure 1**: Map of 654 CPC classes; USPTO data; co-classifications Jaccard normalized; VOSviewer used for classification and visualization. This map can be web-started at
http://www.vosviewer.com/vosviewer.php?map=http://www.leydesdorff.net/cpcmaps/m0.txt&network=http://www.leydesdorff.net/cpcmaps/n0.txt&label_size_variation=0.4&scale=1.15&colored_lines&curved_lines&n_lines=2000 or http://j.tinyurl.com/z9u4nv4 .

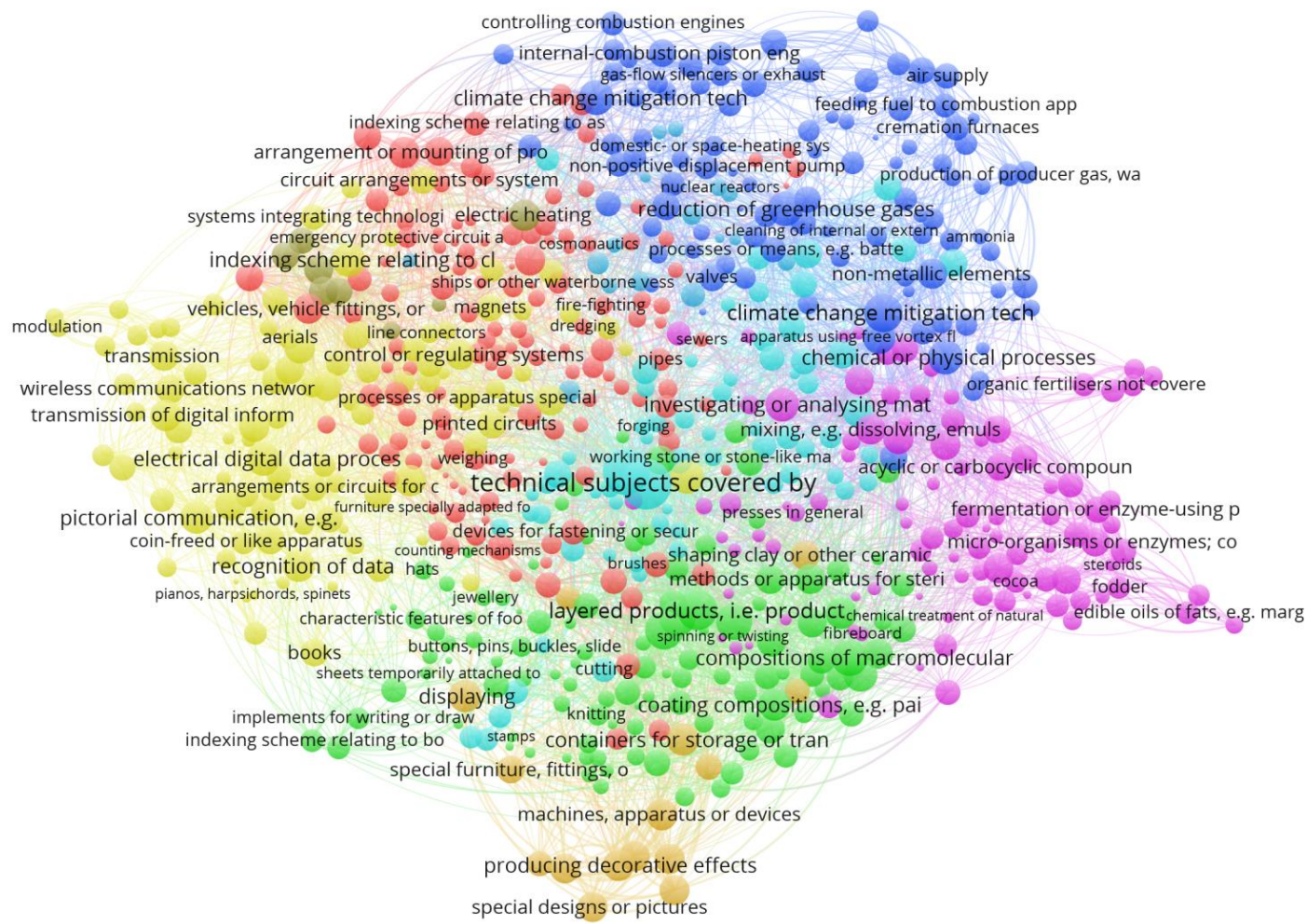

**Figure 2**: Map of 654 CPC classes; USPTO data; co-referencing cosine normalized; VOSviewer used for classification and visualization. This map can be web-started at
http://www.vosviewer.com/vosviewer.php?map=http://www.leydesdorff.net/cpc_cos/m0.txt&network=http://www.leydesdorff.net/cpc_cos/n0.txt&label_size_variation=0.4&scale=1.15&colored_lines&curved_lines&n_lines=5000 or http://j.tinyurl.com/zdbwdn9 .

**Table 1.** Ten most frequently used words in six clusters based on the Jaccard index (stopword-corrected).

| Cluster 1 | Cluster 2 | Cluster 3 | Cluster 4 | Cluster 5 | Cluster 6 |
|---|---|---|---|---|---|
| *material* | vehicle | *material* | system | engine | *material* |
| metal | *device* | compound | apparatus | combustion | composition |
| machine | equipment | treatment | *device* | machine | compound |
| apparatus | apparatus | processes | electric | apparatus | macromolecular |
| printing | arrangement | production | measuring | plant | apparatus |
| tool | building | solid | circuit | steam | article |
| textile | machine | apparatus | control | gases | associated |
| provided | rail | chemical | arrangement | nuclear | coating |
| article | adapted | covered | communication | displacement | covered |
| manufacture | construction | machine | musical | fluid | flat |

**Table 2.** Ten most frequently used words in nine clusters based on cosine-normalization (stopword-corrected).

| Cluster 1 | Cluster 2 | Cluster 3 | Cluster 4 | Cluster 5 | Cluster 6 | Cluster 7 | Cluster 8 | Cluster 9 |
|---|---|---|---|---|---|---|---|---|
| vehicle | *material* | engine | system | *material* | metal | nuclear | cardboard | *device* |
| *device* | apparatus | combustion | apparatus | treatment | *material* | radiation | paper | lighting |
| machine | Printing | machine | *device* | compound | tool | reactor | article | system |
| apparatus | textile | apparatus | measuring | processes | *device* | technique | clay | application |
| door | article | *material* | arrangement | similar | machine | discharge | *material* | associated |
| equipment | machine | production | circuit | apparatus | metallic | explosive | special | electric |
| provided | fabric | fluid | communication | Covered | processes | otherwise | accessories | indexing |
| building | Indexing | gases | control | fertiliser | provided | particle | animal | light |
| engine | relating | heat-exchange | electric | foodstuff | additive | plant | apparatus | relating |
| rail | scheme | solid | instrument | machine | cutting | power | artificial | scheme |

Reading Figure 2 clockwise, one first encounters at the top right (between 12 and two o'clock) a cluster about energy, including labels about technologies for "mitigation of climate change" and "reduction of greenhouse gases." Following the clock, one observes a pink cluster (between 3 and 4 o'clock) with a focus on organic chemistry and enzymes. This cluster is followed (between 5 and 6 o'clock) by a cluster in green with titles referring to macromolecules and paint. Below that a cluster (in light-brown) refers to furniture, etc. To the left (at nine o'clock), we see a cluster in yellow with titles referring to communication. At eleven o'clock, a cluster follows (in orange) with a focus on electrical engineering. A relatively small cluster (in brown, at ten o'clock) between the latter two focuses on light. A cluster colored light-blue is organized along the second diagonal; we would characterize it as "materials and devices." The labels "material" and "device" are italicized in Table 2; they provide the backbone of the map.

One can compare classifications using Cramèr's *V* as a measure of chi-square between zero and one (Table 3). Neither the Jaccard-based clustering in six groups nor the cosine-based in nine groups is statistically independent of the organization of CPC into the nine main classes A to H, and Y ($p< .001$ in both cases).

**Table 3**: Cramèr's *V* among the different classification schemes

|        | Jaccard | Cosine |
|--------|---------|--------|
| Cosine | 0.758   |        |
| CPC-4  | 0.557   | 0.449  |

Table 3 confirms that the two maps are considerably different. However, both are even more different from the CPC-4 classification than from each other. In sum, we will use the cosine-based map (Figure 2) as the basemap for portfolio management. Using CPC.exe, samples downloaded from the USPTO database will be overlaid on this map, similar to the exercise by Kogler *et al*. (2013) where this was performed for U.S. metropolitan areas.

*4.2. Portfolio mapping*

Figures 3a and 3b, for example, show the CPC classes in patents with issue dates in 2016 and inventor addresses in Boston, MA and Eindhoven in the Netherlands, respectively. Surprisingly, the numbers of patents in the USPTO issued in 2016 are of the same order of magnitude for both cities (n = 938 and 1030, respectively.) Eindhoven is a provincial town of approximately 220,000 inhabitants; it has long been the home town of Philips and is nowadays considered a center of technical innovation in the Netherlands. Boston is three times larger than Eindhoven in terms of its population (app. 670,000). Considering that patenting in the USA is not necessarily the first priority for European based companies and inventors, this relative comparability of the patent portfolios was not expected. Rao-Stirling diversity is 0.80 for Boston versus 0.78 for Eindhoven.

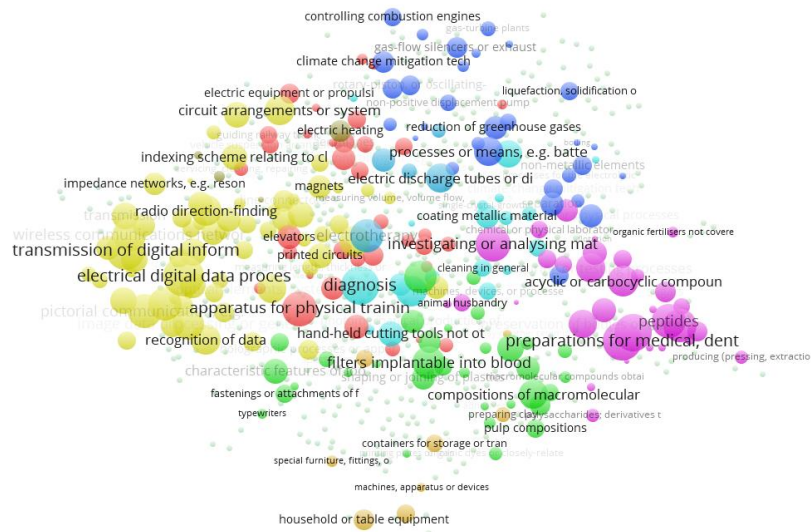 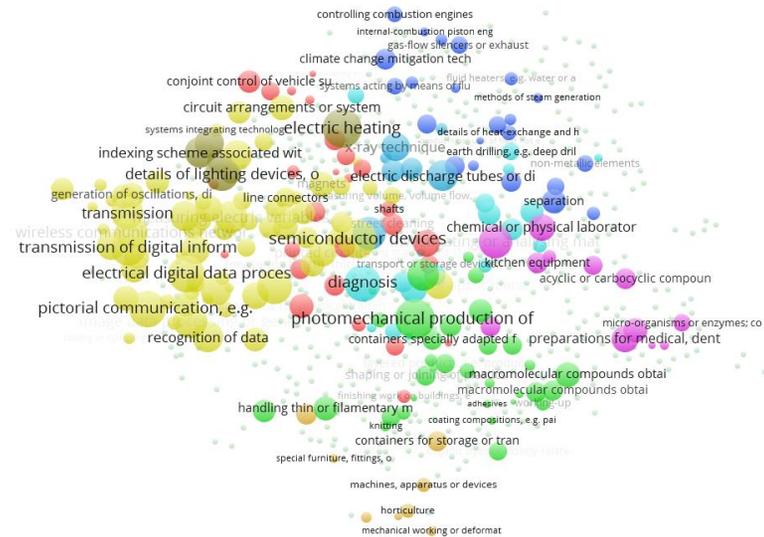

**Figures 3a and 3b**: City portfolios of patents at USPTO for Boston MA (USA; 1030 patents) and Eindhoven (NL; 938 patents) overlaid on the cosine-based patent map of 654 CPC categories at the 4-digit level (Figure 2).
The search strings were "ic/boston and is/ma and isd/2016$$" and "ic/eindhoven and icn/nl and isd/2016$$", respectively.

*3.3. Diversity*

Kogler *et al.* (2017a) used the portfolios granted to inventors in the twenty cities in 2014, but based on IPC-4. We reproduce their results here using issue dates in 2016, CPC classes, and the new map. The objective is to learn more about the operational difference between using IPC-4 and CPC-4: in which respects do the classification schemes make a difference? In Table 4, the cities are listed with the number of patents retrieved in both 2014 and 2016. The data for these 20 cities in 2014 and 2016 are rank-correlated with $\rho = .98$ ($p<.01$). In other words, the retrieval is not incidental.

**Table 4**: Twenty cities in five countries with retrieval for issue dates 2016 and 2014 (Kogler *et al.*, 2017a).

| *China* | 2016 | 2014 | *France* | 2016 | 2014 | *Israel* | 2016 | 2014 | *Netherlands* | 2016 | 2014 | *USA* | 2016 | 2014 |
|---|---|---|---|---|---|---|---|---|---|---|---|---|---|---|
| Beijing | 4043 | 2,122 | Paris | 1233 | 1,336 | Jerusalem | 312 | 283 | Amsterdam | 241 | 253 | Boston | 1030 | 874 |
| Shanghai | 2397 | 1,669 | Marseille | 12 | 13 | TelAviv* | 856 | 876 | Rotterdam | 91 | 102 | Atlanta | 1100 | 1,166 |
| Nanjing | 213 | 192 | Grenoble | 418 | 422 | Haifa | 768 | 776 | Eindhoven | 938 | 884 | Berkeley | 873 | 854 |
| Dalian | 54 | 39 | Toulouse | 301 | 324 | BeerSheva* | 79 | 55 | Wageningen | 45 | 43 | Boulder | 892 | 910 |

\* The search string for Tel-Aviv is: "(ic/tel-aviv or ic/telaviv) and icn/il and isd/2014$$"
\*\* The search string for BeerSheva is: "(ic/beer-sheva or ic/beersheva) and icn/il and isd/2014$$"

**Table 5**: Twenty cities under study ranked by the diversity in their patent portfolios.

| City | Rao's Δ | $^2D^3$ | N |
|---|---|---|---|
| Paris | 0.83 | 5.93 | 1233 |
| Boston | 0.80 | 5.01 | 1030 |
| Rotterdam | 0.80 | 4.89 | 91 |
| Jerusalem | 0.79 | 4.75 | 312 |
| Atlanta | 0.78 | 4.62 | 1100 |
| Eindhoven | 0.78 | 4.62 | 938 |
| Nanjing | 0.78 | 4.61 | 213 |
| Berkeley | 0.78 | 4.53 | 873 |
| Shanghai | 0.78 | 4.49 | 2397 |
| Boulder | 0.78 | 4.48 | 892 |
| Beersheva | 0.78 | 4.46 | 79 |
| Amsterdam | 0.76 | 4.19 | 241 |
| Beijing | 0.71 | 3.44 | 4043 |
| Toulouse | 0.71 | 3.41 | 301 |
| Telaviv | 0.71 | 3.41 | 856 |
| Marseille | 0.70 | 3.31 | 12 |
| Haifa | 0.69 | 3.26 | 768 |
| Grenoble | 0.69 | 3.24 | 418 |
| Dalian | 0.69 | 3.19 | 54 |
| Wageningen | 0.50 | 1.98 | 45 |

Table 5 provides the diversity values for these 20 cities. The number of patents is not correlated to the diversity scores: the Pearson correlation between $N$ and $\Delta$ is 0.16 (and 0.12 between $N$ and $^2D^3$; *n.s.*). Metropolitan cities such as Paris and Boston have the highest scores, whereas patents from Wageningen—the home town of the single agricultural university in the Netherlands—are specific. The results, however, can also be counter-intuitive and thus raise further questions for follow-up research. Rotterdam, for example, scores third on diversity behind Paris and Boston. One can envisage a systematic study of diversity in global or so-called "smart cities," etc., as a perspective for future research.

The values are different from the ones based on IPC-4 in 2014: Shanghai, for example, which was ranked in the first position in 2014 based on IPC, is now rated in 9[th] position once CPC serves as the underlying classification system. The change from IPC to CPC thus makes a difference in this evaluation; the rank-ordering differs significantly between 2014 and 2016 ($\rho =$ .42; *n.s.*).

Using city names, one should be aware that some cities can be more administratively circumscribed than others. For the USA, a classification in terms of metropolitan areas is available. The Core-based Statistical Area (CBSA)[10] of Boston, for example, can be searched in USPTO using the following string: "(ic/(Essex OR Middlesex OR Norfolk OR Plymouth OR Suffolk OR Boston OR Cambridge) AND IS/MA) OR (ic/(Quincy OR Rockingham OR Strafford) AND IS/NH) AND ISD/2016$$". The retrieval is 2,521 as against 2,265 in 2014. However, since there is no CBSA equivalent for the other countries, we abstained in this study from this further elaboration (Grossetti *et al.*, 2014; Maisonobe, Eckert, Grossetti, Jégou, & Milard, 2016).

*4.4.     Portfolio analysis*

A matrix like the one of (20) cities versus (654) CPC-4 classes can be used for various forms of statistical analysis. For example, one can use discriminant analysis given the grouping of cities in countries. Nineteen of the twenty profiles are correctly identified by nation using the CPC classes; the exception is Beersheva, which is predicted as belonging to the French group.

---

[10] A Core Based Statistical Area (CBSA) is a U.S. geographic area defined by the Office of Management and Budget (OMB) that consists of one or more counties (or equivalents) anchored by an urban center of at least 10,000 people plus adjacent counties that are socioeconomically tied to the urban center by commuting.



However, the prediction is considerably improved compared to the one based on IPC, when only 17 or the 20 (85%) were correctly predicted.

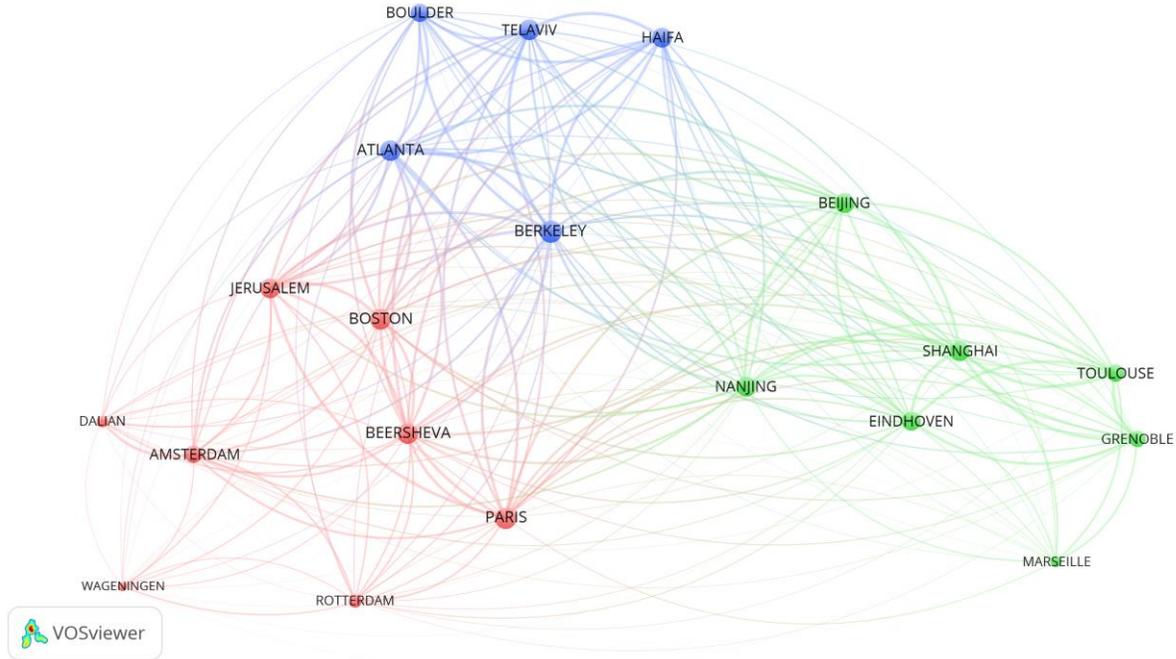

**Figure 4**: Cosine-normalized map of 20 cities in terms of co-occurrences of CPC-4.

Using the cosine values of CPC vectors among the 20 cities, one can generate a local map among the cities differently from the overlay on the global map for each of them (Figure 4). VOSviewer distinguishes three groups. Table 6 shows the three-factor solution of the matrix. (Factor analysis is based on the Pearson correlations. Unlike the cosine, the Pearson correlation normalizes to the mean.)



**Table 6**: Three-factor solution of the matrix of cities versus 654 CPC-4 classes.

**Rotated Component Matrix**[a]

|  | Component 1 | Component 2 | Component 3 |
|---|---|---|---|
| Telaviv | .917 | .113 |  |
| Berkeley | .907 | .245 | .216 |
| Haifa | .876 | .171 |  |
| Atlanta | .869 | .144 | .204 |
| Boulder | .804 |  |  |
| Boston | .799 | .120 | .359 |
| Jerusalem | .754 |  | .406 |
| Beersheva | .709 | .160 | .498 |
| Paris | .667 | .276 | .505 |
| Grenoble |  | .960 |  |
| Toulouse | .102 | .956 |  |
| Shanghai | .332 | .908 |  |
| Eindhoven | .340 | .770 |  |
| Beijing | .572 | .583 | .103 |
| Nanjing | .538 | .548 | .213 |
| Marseille |  | .496 |  |
| Amsterdam | .287 |  | .837 |
| Dalian |  |  | .706 |
| Rotterdam | .177 |  | .469 |
| Wageningen |  |  | .463 |

Extraction Method: Principal Component Analysis.
Rotation Method: Varimax with Kaiser Normalization.
a. Rotation converged in 5 iterations.

This three-factor solution explains 69.79% of the variance. This solution is only weakly related to the corresponding three-factor matrix made on the basis of 2014 data and IPC ($r = .31$; $p <= .05$). Major cities are grouped as factor 1; factor 2 is related to this first factor by Chinese cities (Beijing, Shanghai, and Nanjing), which share a pattern also with technologically oriented cities such as Grenoble, Toulouse, and Eindhoven. Factor 3 is set apart with other factor loadings of



three Dutch cities (Amsterdam, Rotterdam, and Wageningen) and Dalian in China. Why the grouping is in some respects different between Figure 4 and Table 6 require further analysis beyond the scope of this study.

*4.3.Visual comparisons between two portfolios*

The routine CPC2.exe (available at http://www.leydesdorff.net/cpc_cos/portfolio) allows for the further extension of comparing two portfolios with each other in a single map. (For instructions see the Appendix or at http://www.leydesdorff.net/cpc_cos/portfolio.)

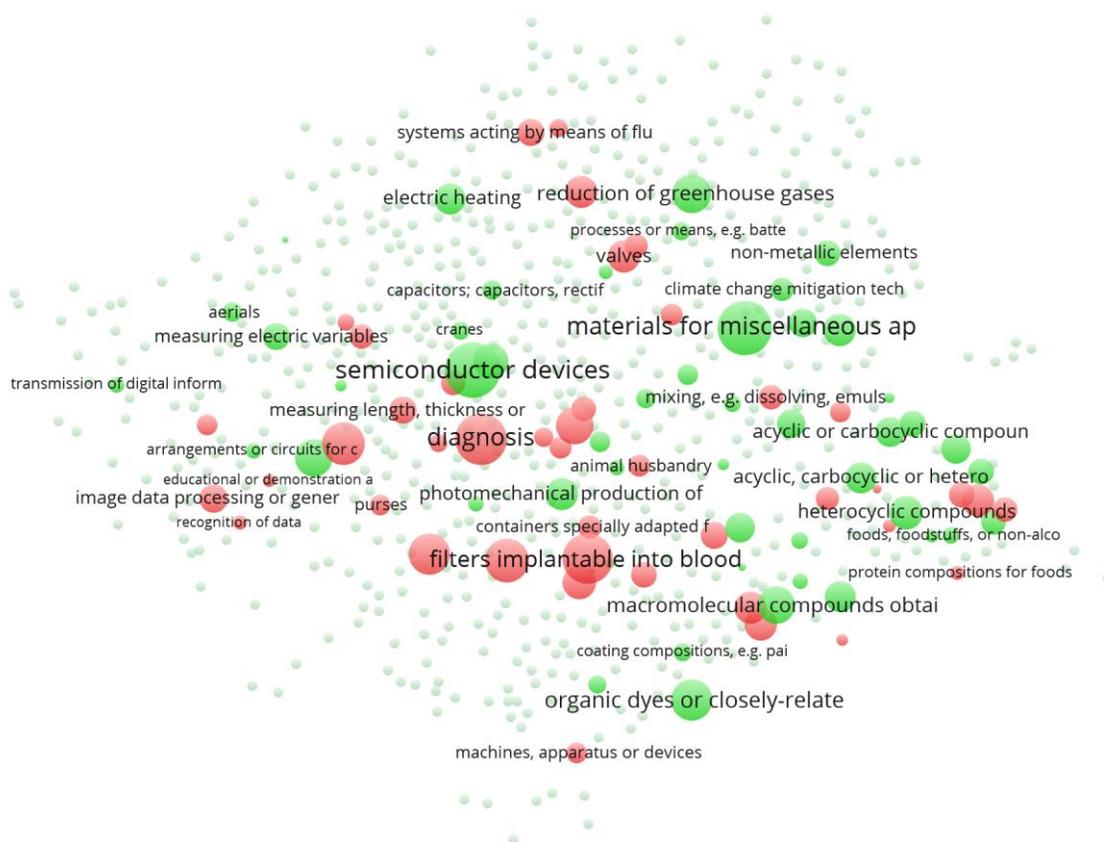

**Figure 5**: Comparison between 276 patents granted to Novartis *vs*. 350 patents granted to Merck



Sharpe and Dome in 2016. This map can be web-started at http://www.vosviewer.com/vosviewer.php?map=http://www.leydesdorff.net/cpc_cos/portfolio/fig9b.txt&label_size_variation=0.25&scale=1.0 or http://j.tinyurl.com/hwz6275

Applying the same procedure as above for cities, but now using firm names the retrieval was 276 patents granted at USPTO in 2016 using, for example, the search string "(an/Novartis or aanm/Novartis) and isd/2016$$". "AN" and "AANM" are fields code for the names of assignees and applicants, respectively, in the advanced search engine of USPTO. Note that the applicant is not always the assignee of a patent; the applicant may also be the inventor, whereas the assignee is the owner of the patent.

In Figure 5, the resulting map is compared with a similar one using the 350 patents granted to MSD in 2016. The routine CPC2.exe provides a file vos2.txt that can be used as input to VOSviewer (and corresponding files for Pajek). The red-colored nodes on this map indicate the CPC-4 classes in which Novartis has more patents than MSD, and the green-colored ones the classes with more patents of MSD than Novartis. Novartis, for example, has a more pronounced portfolio in the region in the figure labeled "filters implantable into blood"; the portfolio of MSD is stronger in the area indicated around "semiconductor devices".

**Table 7**: Diversity in the patents granted to Novartis and MSD at USPTO in 2016.

|  | *Rao's Δ* | *$^2D^3$* | *N* |
|---|---|---|---|
| Novartis | 0.53 | 2.13 | 276 |
| MSD | 0.60 | 2.52 | 350 |

Table 7 shows the diversity measures for the two companies. As might be expected, the values for diversity are much lower in the case of specific industries or firms than in the case of cities (see Table 5).

## 4. Discussion and concluding remarks

Indicators for measuring the distance or relatedness between distinct sets of technological knowledge categories are of interest in a number of disciplines; for example, management studies (Almeida, 1996; Makri *et al.*, 2010), economics (Jaffe, 1986; Teece *et al*., 1994), and regional science (Fischer *et al*., 2006; Quatraro, 2010; Boschma *et al*., 2013). The strength of the relationships between unique sets of technological expertise can inform us about the level of specialization and coherence amongst firms, regions, and/or countries. This in turn has been linked to the level of productivity of these units (Nesta and Saviotti, 2005; Kogler *et al*., 2013), and can also serve as an indicator for potential diversification and innovation opportunities as well as limitations (Colombelli *et al.*, 2014; Feldman *et al*., 2015; Kogler *et al*., 2017b).

One of our objectives has been to show how sensitive the maps and overlays are to choices of parameters. The maps are not natural, like geographic maps, but remain constructs that originate from discursive reasoning. Consequently, one cannot infer on the basis of visual inspection that one representation is better than another; but one needs analytical arguments for the choices. For this reason, it was urgent to take Yan & Luo's (2017) critique seriously and redo the maps. As we have shown, the results based on CPC-4 are roughly similar, but different in important respects from IPC-based maps. In our opinion, the ostensable similarity is to a large extent due to the use of similar methods; the empirical results are in important respects significantly different.

We did not follow Yan & Luo's preference for the Jaccard index, but used the cosine. The Jaccard index is relational and binary, whereas the cosine is a numerical and a positional measure (Burt, 1982; Leydesdorff, 2014). Portfolios projected on a global map can be considered as positions. We suggested to study relations using a local map and provided a tool for this purpose.

The global maps were not interpretable when using the Jaccard index (Figure 1), but they were interpretable on the basis of the cosine (Figure 2). However, we agree on using the basic (2-mode) matrix of cited patents versus citing aggregates into CPC classes. Using this disaggregated level can be considered as an important improvement on previous maps. The input data is more finely-grained by orders of magnitude.

The multi-variate analysis confirmed the improvement. The prediction based on discriminant analysis was improved from 85 to 95%. The (varimax) rotation of the eigenvectors provided us with additional and meaningful insights in the comparisons among cities. The positions of the major Chinese cities among cities in the global domain (factor 1) and cities with strong engineering (factor 2) are suggested in the cosine-based map.

Visualization is a strong instrument because maps can be provided with an interpretation more easily than numerical statistics. However, the latter are needed in the background for making analytical arguments. A direct comparison between two units (e.g., competing firms) using a map, however, can provide a first orientation to their differences in terms of strengths and weaknesses.

**Acknowledgement**
We are grateful to Jordan Comins and two anonymous referees for comments on a previous draft. BY acknowledges support by the Academic Research Fund Tier 2 of the Singapore Ministry of Education. DK acknowledges funding by the European Research Council, grant nr. 715631.

**Appendix**

1. **Preparing input files**
   a. Download the following files from http://www.leydesdorff.net/cpc_cos/portfolio (or https://leydesdorff.github.io/cpc/portfolio) into a single folder on your hard disk:
      - cpc.exe;
      - cpc.dbf (with basic information about the classes);
      - uspto1.exe (needed for the downloading of USPTO patents);
      - cos_cpc.dbf (needed for the computation of distances on the map);
   b. Run cpc.exe.

2. **Options within cpc.exe**
   a. The program asks for a short name (≤ 10 characters) in each run. This name will be used as the variable label in later parts of the routine;
   b. The first option is to download the patents from USPTO at http://patft.uspto.gov/netahtml/PTO/search-adv.htm ; detailed instructions for the downloading can be found at http://www.leydesdorff.net/ipcmaps;
   c. USPTO has a maximum of 1000 records at a time, but one is allowed to follow-up batches; after the download is completed, save the files in another folder or as a zip file;

3. **The incremental construction of the files matrix.dbf and rao.dbf**
   a. After each run, a column variable is added to the (local) file matrix.dbf containing the distribution of the 654 CPC classes in the document set under study. If the file matrix.dbf is absent, it is generated *de novo* and the current run is considered as generating the first variable; matrix.dbf can be read by Excel, SPSS, etc., for further (statistical) analysis;
   b. Similarly, a row variable is added after each run to the file rao.dbf containing diversity measures (explained in the article) as variables. This file is also generated *de novo* if previously absent. Distances are based on $[1 - \cos(x,y)]$ for each two distributions **x** and **y** of aggregated citation at the level of CPC-4 classes;
   c. The routine cpc2cos.exe reads the file matrix.dbf and produces cosine.net and coocc.dat as (normalized) co-occurrence matrices that can be used in network analysis and visualization programs such as Pajek or UCInet.

4. **Output files in each run**
   a. The file "vos.txt" can be read by VOSviewer for mapping the portfolio under study at the four-digit level of CPC; the distances and colors (corresponding to clusters) in the maps are based on the base-map provided in Figure 2 above;
   b. The files cpc.vec and cpc.cls can be used as a vector and cluster files in the Pajek file provided at http://www.leydesdorff.net/cpc_cos/ . This allows for layouts other than VOSviewer and for more detailed network analysis and statistics. The file cpc.cls is a so-called cluster file which can be used in Pajek, among other things, for the extraction of partitions.
c. The various fields in the USPTO records are organized in a series of databases that can be related (e.g., in MS Access) using the field "nr".

## 5. Visual comparison among portfolios (using cpc2.exe)

One can compare two portfolios (as in Figure 5 above) using cpc2.exe (available at http://www.leydesdorff.net/cpc_cos/portfolio/cpc2.exe ).
a. One first runs cpc.exe for the one set (e.g., city1);
b. Replace the downloaded patents (p1.htm, p2.htm, etc.) with the set for the second unit (e.g., city2) and run cpc2.exe;
c. The file vos2.txt generated is an input file to VOSviewer. The red-colored nodes indicate the CPC-4 classes in which the first unit is stronger than the second; the green-colored nodes indicate the relative strength of the second set;
d. The files cpc2.vec and cpc2.cls provide the corresponding input files for Pajek.